\documentclass[aps,pre,floats,twocolumn,showpacs]{revtex4-1} 

\usepackage{graphicx,epsfig}
\usepackage{graphics,bm,epic,eepic,float}
\usepackage{amssymb,amsmath,multirow,rotate,color}
\usepackage{subfigure}
\usepackage{color}
\usepackage{times}

\def\Erdos{Erd\"os}

\begin{document}

\title{Impact of network structure on a model of diffusion and
  competitive interaction}

\author{V. Nicosia$^{1,2}$, F. Bagnoli$^3$, V. Latora$^{1,2}$}

\affiliation{ $^{1}$~Dipartimento di Fisica e Astronomia, Universit\`a
  di Catania and INFN, Via S. Sofia, 64, 95123 Catania, Italy}
\affiliation{ $^{2}$~Laboratorio sui Sistemi Complessi, Scuola
  Superiore di Catania, Via San Nullo 5/i, 95123 Catania, Italy}
\affiliation{ $^{3}$~Dipartimento di Energetica, Universit\`a di
  Firenze, and CSDC and INFN, sez. Firenze, Firenze, Italy}

\begin{abstract} 
  We consider a model in which agents of different species move over a
  complex network, are subject to reproduction and compete for
  resources. The complementary roles of competition and diffusion
  produce a variety of fixed points, whose stability depends on the
  structure of the underlying complex network. The survival and death
  of species is influenced by the network degree distribution,
  clustering, degree-degree correlations and community structures. We
  found that the invasion of all the nodes by just one species is
  possible only in \Erdos--Renyi and regular graphs, while networks
  with scale--free degree distribution, as those observed in real
  social, biological and technological systems, guarantee the
  co--existence of different species and therefore help enhancing
  species diversity.
\end{abstract}


\maketitle 

In the last few years, complex networks have been the subject of an
increasingly large interest in the scientific
community~\cite{rev1,rev2,rev3}.  This is due to: {\em i)} the wide
variety of complex systems that can be described as graphs with a
complex topology; {\em ii)} the recent observation that various
dynamical processes taking place on networks, such as
epidemics~\cite{epidemics}, random walks~\cite{noh04,brw},
synchronisation~\cite{arenas} and self--organised
criticality~\cite{moreno,JensenBook}, can be affected by the
underlying network structure.  Concerning social interactions, many
models have been proposed in the last decades to study evolution of
relationships, culture segregation, opinion formation and propagation
of new ideas by means of majority rules, melting or mixing, and have
also been studied on complex
topologies~\cite{Jensen01,Jensen02,Schelling,Axelrod,Castellano}.
Particularly important contributions to the understanding of social
dynamics taking place on networks have been provided by evolutionary
game theory~\cite{hs,nowak}. Usually, when an evolutionary game is
studied on a graph, each individual is associated to a node of the
graph, and the social relationships are represented by the
links. Individuals interact with their neighbours on the graph by
playing various evolutionary games, and different collective
behaviours emerge, such as global cooperation or selfishness,
according to the structure of the underlying network of
relationships~\cite{santos,game1}.  These models on complex networks
fail to catch one of the most important characteristic of real
evolutionary systems, namely the possibility for the individuals to
move through a complex environment and to interact with a
neighbourhood which changes over time~\cite{tang}. Some recent works
have proposed extensions of evolutionary game theory to moving
agents~\cite{Helbing2,Meloni}. In these latter models, the individuals
move over a continuous space or a discrete lattice, and play games
with other individuals in their spatial neighbourhood.  However, the
hypothesis of a homogeneous and continuous space is too simplistic,
and does not correspond to the structure of real social and
technological networks~\cite{rev1,rev2,rev3,airplane}.  Metapopulation
models with heterogeneous connectivity patterns, which incorporate
mobility over the nodes, local interaction at the nodes, and a complex
network structure, have been recently proposed only in specific
contexts such as epidemic spreading ~\cite{eubank} or chemical
reactions ~\cite{Colizza2}. In this Letter we propose and study a
simple and general model of evolution of species over a complex
network. In the model, each species can represent either a biotype or
a language, a culture or even a consumer product.  The species compete
for space or resources, represented by the nodes of the network.  For
instance, if the species are consumer products, each node is a
potential user and the species move from node to node through the
network of social relationships among users, competing to be adopted
by as many users as possible. Instead, if we imagine each species as a
different biotype, the complex network represents the connections
among spatial environments, and species compete for food or energy.
The agents of different species move over the graph by diffusion and
their interaction at the nodes is modelled by means of competition and
replication rules.  The aim of the present work is to study the
combined effects of \textit{diffusion} over a complex topology and of
\textit{competitive selection} at nodes. Notice that whenever we refer
to \textit{selection} in the following, we always intend
\textit{competitive selection}, i.e. competition for resources among
the different species at the same node. As we will show in the
following, the model, although extremely simple, is rich enough to
exhibit a large variety of patterns over time and a final distribution
of the species that is intrinsically connected with the structure of
the network.

Let us consider a connected graph with $N$ nodes and $K$ links,
described by an adjacency matrix $A=\{ a_{ij} \}$, and a population of
$N_s$ different species, labelled by the index $\alpha=1,2,\ldots,N_s$.
Each node of the graph is an environmental niche which can host
individuals of one or more species, and the links represent
connections between niches. At every single node of the graph, there
is room for each of the $N_s$ species. We denote with
$P_i^{\alpha}(t)$ the relative abundance of the $\alpha$-th species at
node $i$ at time $t$. Such abundances are normalised so that all nodes
have the same capacity, i.e.  $\sum_{\alpha} P_i^{\alpha}(t) =1,
\forall i, t$.  We denote with $P^{\alpha}(t) = \sum_i
P_i^{\alpha}(t)$ the overall abundance of species $\alpha$ in the
network, so that $\sum_{\alpha} P^{\alpha}(t) = N$.  At each time
step, the model considers two different processes, namely diffusion
and competition. 
In the {\em diffusion process}, a fraction $p$ ($0 \leqslant p
\leqslant 1$) of the individuals which are at a given node $j$ move to
one of the first neighbours of $j$, let us say $i$, with a uniform
probability: $ \frac{a_{ji}}{\sum_l a_{jl}}$. The remaining fraction
$1-p$ stays at node $j$.  After the diffusion process, the {\em
  selection process} takes place, which normalises the number of
individuals at each node in order to guarantee that $\sum_\alpha
P_i^{\alpha}(t) = 1, \forall i,t$. The survival and death of
individuals at each node is governed by a generalised replicator
dynamics~\cite{hs,nowak}.  In its simplest version, which takes into
account an ecosystem with only two species, say $X$ and $Y$, the
equations of the replicator dynamics read:
\begin{equation}
  \begin{split}
    x(t+1) = \dfrac{f[x(t)]}{\phi} x(t), &\quad y(t+1) =
    \dfrac{g[y(t)]}{\phi} y(t)
  \end{split}. 
  \label{eq:discr_actual}
\end{equation}
where $x(t)$ and $y(t)$ denote the percentages of individuals of
species $X$ and $Y$ at time $t$, respectively, while $f[x(t)]$ and
$g[y(t)]$ are two functions which measure the \textit{fitness} of each
of the two species. The interaction between $X$ and $Y$ is ruled by
the quantity $\phi$, which plays the role of an environmental limit
and is fixed to ensure the normalisation $x(t)+y(t)=1 ~ \forall
t$. This gives $\phi = xf(x) + y g(y)$, so that $\phi=\phi(x,y)$ is
the average fitness of the population. The meaning of
Eq.~(\ref{eq:discr_actual}) is the following: when the only constraint
imposed to the species evolution is the environmental limit $\phi$,
their relative abundance at the next time step will increase or
decrease according to their fitness. As for the fitness functions
$f(x)$ and $g(y)$ we consider the general case~\cite{nowak} $f(x) =
b_x x^{\gamma - 1}$ and $g(y) = b_y y^{\gamma - 1}$ where $b_x>0$,
$b_y>0$, and the exponent $\gamma$ is a real number that can be varied
to tune the dependence of the fitness function of a species on its
abundance.  The fixed points $(x^*,y^*)$ of
Eq.~(\ref{eq:discr_actual}), and their stability, depend on the value
of $\gamma$. We distinguish three cases: $\gamma$ smaller, equal or
larger than $1$.
For $\gamma<1$ there are two unstable fixed points $(1,0)$ and
$(0,1)$, and one stable fixed point
$x^*=(1+(b_x/b_y)^{\frac{1}{\gamma-1}})^{-1}$, $y^*=1-x^*$. The two
species $X$ and $Y$ will coexist despite their initial relative
abundances and fitness. This case is called \textit{survival for all}.
If $\gamma=1$, there are only two fixed points whose stability depends
on the respective values of $b_x$ and $b_y$.  When $b_x>b_y$ then
$(1,0)$ is stable while $(0,1)$ is unstable, conversely when
$b_x<b_y$, $(0,1)$ is a stable equilibrium while $(1,0)$ in
unstable. Independently of the initial distributions of the two
species, if $b_x > b_y$ then $x$ will eventually overcome $y$ until
all individuals of species $y$ are extinct. This case is called
\textit{survival of the fittest}.
Finally, for $\gamma>1$ the third fixed point is unstable, while
$(1,0)$ and $(0,1)$ are both stable. In particular if the initial
condition $x(0)$ is such that $x(0)>x^{*}$, then $x(t)$ will
eventually overcome $y(t)$, independently of the value of $b_x$ and
$b_y$, while $y(t)$ will overcome $x(t)$ if $x(0) < x^{*}$ (notice
that $x^{*}=1/2$ when $b_x=b_y$). This super-exponential growth always
guarantees the survival (and reproduction) of the most abundant
species, so that the case $\gamma>1$ is usually called
\textit{survival of the first}.  In order to implement a strong
competition among species at a node, in the following we always
consider $\gamma>1$. The replicator dynamic can be easily extended to
$N_s$ different species. A super-exponential growth is predicted for
the $N_s$-dimensional replicator equations when $\gamma>1$, and all
the corners of the $N_{s}$-dimensional simplex are stable fixed
points. We consider a fitness function of the form $f(P_i^{\alpha}) =
b_i^{\alpha} (P_i^{\alpha})^{\gamma - 1}$, where $\gamma > 1 $ and,
without any lack of generality, we set $b_i^{\alpha} = 1$ $\forall i$
and $\forall \alpha$. In fact, when $\gamma >1$ only the most abundant
species will survive, despite the relative values of $b_i^{\alpha}$
(\textit{survival of the first}). The final model consists of the
following equations:
\begin{eqnarray}
    Q_i^{\alpha}(t) & = &  (1-p) P_i^{\alpha}(t) + p \sum_j \dfrac{
      P_j^{\alpha}(t) a_{ji} }{\sum_l a_{jl}} \label{eq:eq1} \\
    P_i^{\alpha}(t+1) & = & \dfrac{ [Q_i^{\alpha}(t)]^{\gamma}}{
      \sum_{\beta} [Q_i^{\beta}(t)]^{\gamma}} \label{eq:eq2} 
\end{eqnarray}
where $i=1,\ldots,N$. Eq.~(\ref{eq:eq1}) accounts for the diffusion
process while Eq.~(\ref{eq:eq2}) accounts for the selection, with $0
\leqslant p \leqslant 1$ and $\gamma>1$ being the two control
parameters of the model. The quantity $Q_i^{\alpha}$ represents the
local abundance of species $\alpha$ at node $i$ before selection takes
place, while $P_i^{\alpha}$ is the local abundance of species $\alpha$
at node $i$ after selection. The dynamics of the model finally
converges to a stationary state with a fixed number of surviving
species. We have found that this number can vary from $1$ up to $N_s$,
according to the values of the two parameters $p$ and $\gamma$. Notice
that a species $\alpha$ can be considered extinct when $P^{\alpha}(t)
= \sum_i P_i^{\alpha}(t) < \frac{1}{N_{s}(t)N}$, where $N_{s}(t)$ is
the number of species still present on the network at time $t$.  This
comes directly from the observation that when $\gamma > 1$ a species
can grow and exponentially reproduce on a node if and only if it is
the most abundant species on that node. If at time $t$ the overall
abundance of a species is lower than $\frac{1}{N_{s}(t)N}$, then it
cannot be the most abundant species at any node, and will eventually
disappear. A very interesting feature of the model is that the number
of species at equilibrium depends, as expected, on the diffusion
(parameter $p$) and on the strength of interaction (the $\gamma$
exponent), but it is also heavily affected by the topology of the
underlying network. An effective visual representation of the
depencence of the dynamics on the network structure is the {\em phase
  diagram} which reports the number of surviving species as a function
of $p$ and $\gamma$. In fact, the shape of the phase diagram seems to
be tightly connected with the topological structure of the network.
We first show how to derive analytically some information on the
number of surviving species in the simple case of random regular
graphs. The fixed points of Eq.~(\ref{eq:eq2}) and their stability can
be studied analytically in a mean--field approximation. In the
mean--field the adjacency matrix of the graph is expressed in terms of
the probabilty $a_{ij}=\frac{k_ik_j}{2K}$ of having the edge $a_{ij}$
between nodes $i$ and $j$ if $k_i$ and $k_j$ are, respectively, the
degree of node $i$ and node $j$, and $K$ is the number of edges in the
graph~\cite{bianconi}. Using the mean--field approximation for
$a_{ij}$ in Eq.~(\ref{eq:eq1}) and substituting back in
Eq.~(\ref{eq:eq2}), we obtain the following time evolution for the
occupation probabilities:
\begin{equation}
  P_i^{\alpha}(t+1) = \dfrac{[P^{\alpha}(t)]^{\gamma}\left[
      (1-p)\frac{P_i^{\alpha}(t)}{P^{\alpha}(t)} + p
      \frac{k_i}{2K}\right]^{\gamma}}{\sum_{\beta}[P^{\beta}(t)]^{\gamma}\left[
      (1-p)\frac{P_i^{\beta}(t)}{P^{\beta}(t)} + p
      \frac{k_i}{2K}\right]^{\gamma}}\\
  \label{eq:eq_meanfield_P}
\end{equation}
We can therefore look for the fixed points: $P^{\alpha}(t+1) =
P^{\alpha}(t) ~\forall \alpha$, and check for their linear stability
to small perturbations.  In the following we consider the case
$N_S=N$, i.e. a number of species equal to the number of nodes. It is
easy to prove that in this case state $S_1 \equiv \{P^{\alpha} = 1,
\forall \alpha\}$ is a stable fixed point for $p=1$, and that state
$S_2 \equiv \{P^{\overline{\alpha}} = N, P^{\alpha}= 0~ \forall
\alpha\neq\overline{\alpha}\}$ is a stable fixed point $\forall p$.
In general, finding all the fixed points of
Eqs.~(\ref{eq:eq_meanfield_P}), for any value of $p$ and $\gamma$, is
not an easy task because of the dependencies of the equations on the
node degrees. A drastic simplification is obtained if we make the
assumption that all the nodes have the same degree $k_i = k =
\frac{2K}{N}$, i.e. when the graph is regular. It is easy to verify
that for regular graphs in the mean--field approximation, the state
$S_3 \equiv \{P^{\alpha} = 1, \forall\alpha; P_i^{\alpha} =
\delta_{i\alpha}\}$ is a fixed point for all values of $p$ and
$\gamma$.  Notice that in state $S_3$ each node contains individuals
of only one species, and each species $\alpha$ is present only on one
node. The edge of the stability region for state $S_3$ is given by
equation:
\begin{equation}
  c \left[\dfrac{x}{(N-1)ay + bx} + \dfrac{N-1}{(N-2)a + b + az}\right] - 1 = 0\\
  \label{eq:s3_stability}
\end{equation}
where
\begin{align*}
  a &=(1+\frac{\varepsilon}{N-1})^{\gamma}&, b &=(1-\varepsilon)^{\gamma}\\
  c &=(1-\varepsilon)^{\gamma - 1}&, x &=\left[\frac{N(1-p)}{p} + 1\right]^{\gamma}\\
  y &=\left[\frac{N(1-p)\frac{\varepsilon}{N-1}}{p(1+\frac{\varepsilon}{N-1})}
    + 1\right]^{\gamma} &, z &=\frac{N(1-p)}{p(1+\frac{\varepsilon}{N-1})} + 1
\end{align*}
and $\varepsilon$ is a small perturbation of the fixed point
$S_3$. Eq.~(\ref{eq:s3_stability}) is obtained by imposing that $S_3$
is a fixed point, i.e. that relative species abundances on the whole
graph remain constant over time, and then performing a small
perturbation on the abundance of just one species.  In particular, we
imagine that one of the species decreases its abundance by a small
amount $\varepsilon$, and that this amount is uniformly redistributed
to the other $N-1$ remaining species, in order to guarantee that the
total amount of individuals on the network remains constant. Notice
that Eq.(\ref{eq:s3_stability}) depends only on $p$, $\gamma$ and $N$,
and does not depend on the number of links $K$, since in the
mean--field approximation each node is connected to all the other
nodes.

\begin{figure}[!htbp]
  \centering
  \includegraphics[scale=0.28]{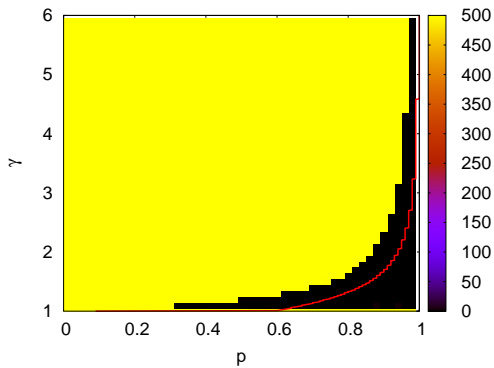}
  \caption{Final number of species $N_s$ on a random regular graph
    with $N=500$ nodes and $k=100$ (averaged over 200 realizations),
    as a function of the two parameters $p$ and $\gamma$.  The yellow
    area corresponds to $500$ surviving species, while black area
    indicates only one surviving species. The red line marks the edge
    of stability for $S_3$ in the mean-field approximation.}
  \label{fig:rnd_reg_mean_field}
\end{figure}
The red line drawn in Fig.~(\ref{fig:rnd_reg_mean_field}) is the
numerical solution of Eq.~(\ref{eq:s3_stability}) for different values
of $(p, \gamma)$ and for $\varepsilon \rightarrow 0$ on a regular
network with $N=500$ nodes. The state $S_3$ with $N$ surviving species
is unstable for values of $(p,\gamma)$ below the red line, while it is
stable for $(p, \gamma)$ above the red line. To check the validity of
the mean--field approximation, we simulated the dynamics of
Eq.~(\ref{eq:eq1}) over a regular random graph with $N=500$ nodes, and
a large average degree $k=100$, for different values of $(p,\gamma)$,
initialising the system in state $S_3$. In
Fig.~(\ref{fig:rnd_reg_mean_field}) we plot as a colour map the number
of species surviving at equilibrium for different values of $p$ and
$\gamma$. Yellow regions correspond to $500$ surviving species, while
black regions correspond to only one surviving species. Notice that
the agreement with the theory is very good: the yellow area
approximately coincides with the region where $S_3$ is stable, while
the black zone corresponds to the region where $S_3$ is
unstable. Notice also that the transition is very sharp and well
defined in the $(p,\gamma)$ plane, meaning that the system suddenly
moves from a state where $N_s=N$ to a state where $N_s=1$.
 \begin{figure*}[!htbp]
   \subfigure[]{
     \includegraphics[scale=0.22]{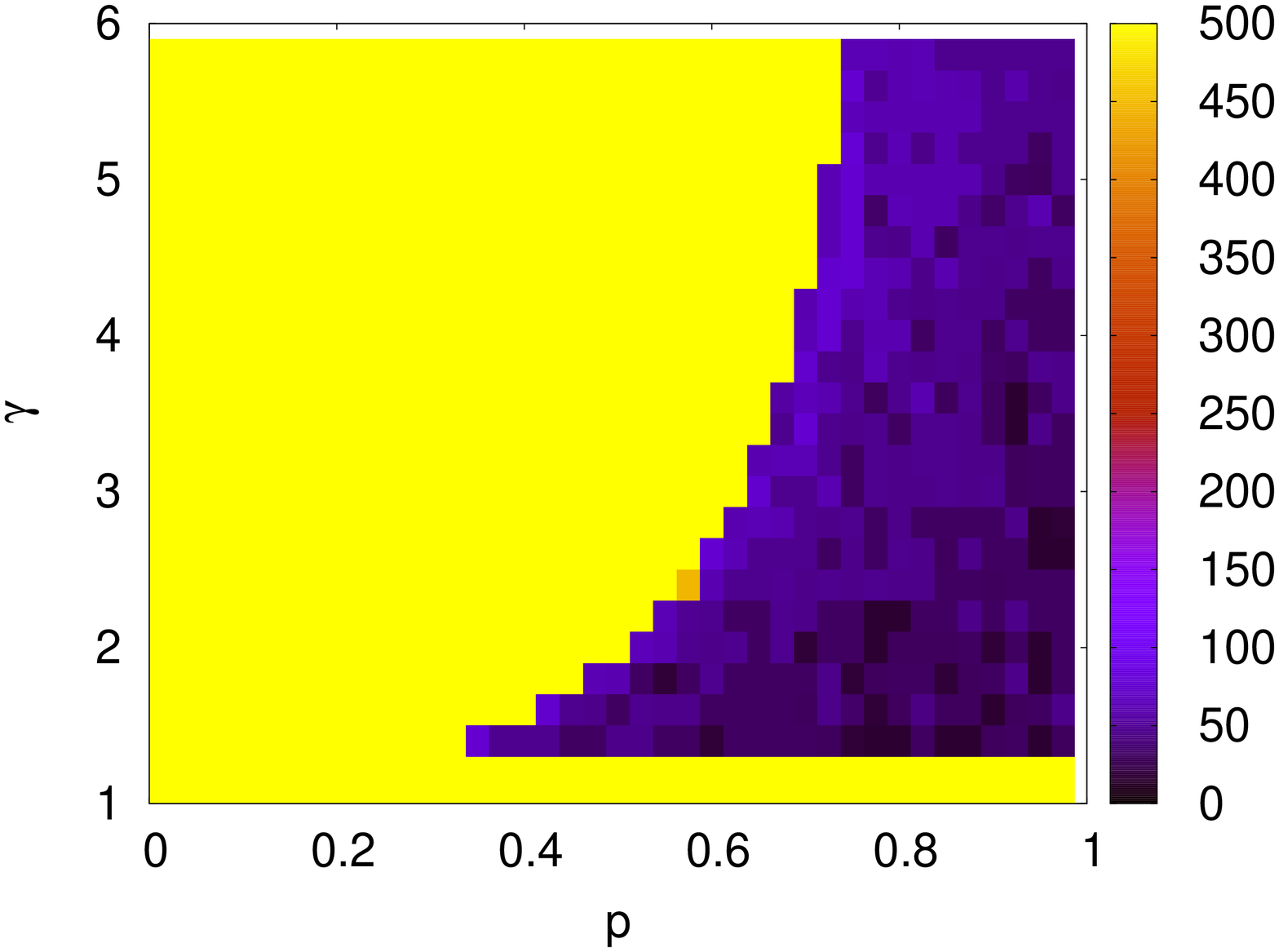}
     \label{fig1a}
   }
   \subfigure[]{
     \includegraphics[scale=0.22]{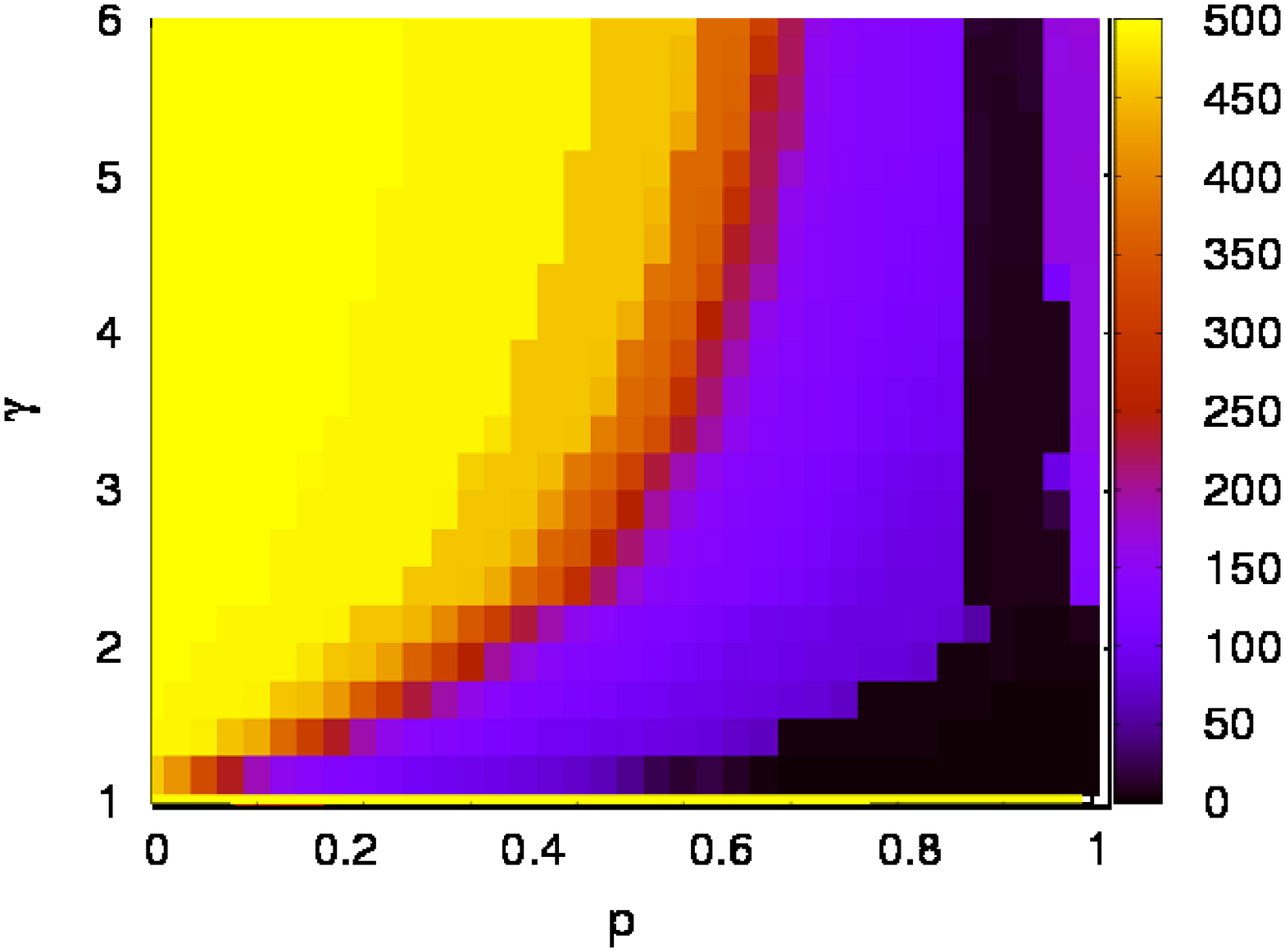}
     \label{fig1b}
   }
   \subfigure[]{
     \includegraphics[scale=0.22]{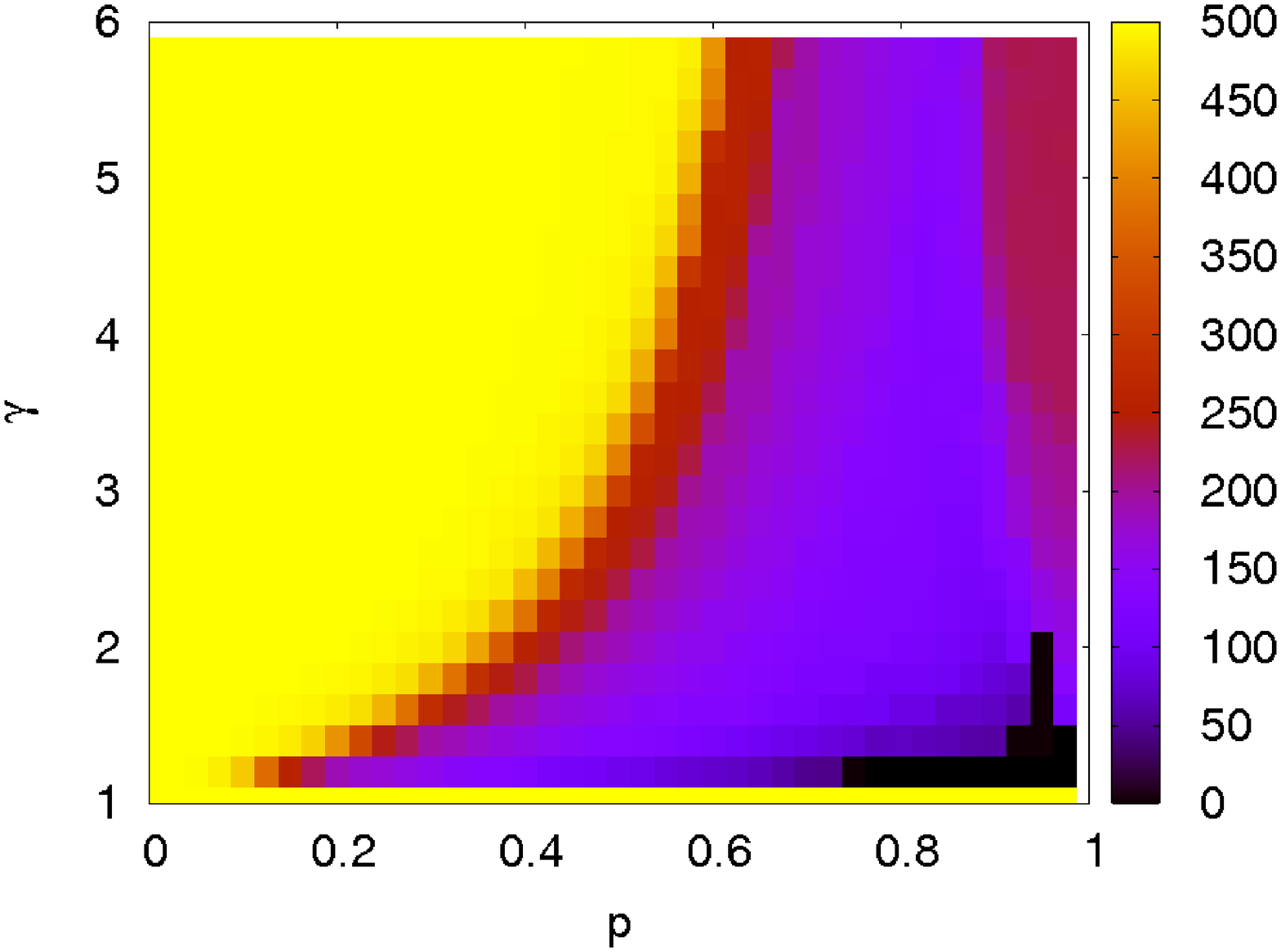}
     \label{fig1c}
   }
  \caption{Number of species $N_s$ on the graph at equilibrium as a
    function of the two parameters $p$ and $\gamma$.  Three classes of
    graphs with $N=500$ nodes and $ \langle k \rangle = 6$ were
    considered: {\em a)} regular ring lattice, {\em b)} \Erdos--Renyi
    random graph, {\em c)} scale--free graph.}
  \label{fig1}
\end{figure*}
To explore the dependence of the dynamics on the degree distribution
of the network, we have simulated Eq.~(\ref{eq:eq1}) with initial
state $S_3$ over three different networks, namely regular lattices
(RL) with periodic boundary conditions, \Erdos--Renyi (ER) random
graphs, and scale--free (SF) graphs with degree distribution $P(K)
\sim k^{-3}$~\cite{rev1,rev2,rev3}. All the graphs have been created
with the same number of nodes ($N=500$) and the same number of links
($\langle k \rangle \simeq 6$). The phase diagrams in
Fig.~(\ref{fig1}) show the number of species $N_s$ remaining at
equilibrium on the three classes of graphs as a function of the two
parameters $p$ and $\gamma$. We observed that a stationary value of
$N_s$ is reached on these graphs after no more than $200$ iterations
of Eq.~(\ref{eq:eq1}), and we have checked that this value does not
change for at least $20000$ iterations. Diagrams for ER and SF graphs
are obtained as an average over 200 realizations, even if the
fluctuations from one realization to another are very small.  We
notice that both $p$ and $\gamma$ play an important role on the final
number of species remaining at equilibrium. For small $p$ and large
$\gamma$ all the species survive, with each species remaining in its
starting node. In fact, when $p$ is close to $0$, only a small amount
of individuals leave their starting nodes and die almost immediately
after they arrive at neighbouring nodes due to selection. This
corresponds to the large yellow area ($N_s=500$) present in all the
phase diagrams reported in Fig.~(\ref{fig1}). As the diffusion
probability $p$ increases, a stronger selection (larger value of
$\gamma$) is needed to prevent the invasion and let all the $N_s=500$
species survive.  Despite some similarities for small $p$ and large
$\gamma$, the three graphs exhibit different behaviour when diffusion
and selection are such that some species can invade neighbouring
nodes, and some other species eventually disappear. According to the
values of $p$ and $\gamma$, the combination of diffusion and
competition determines stationary solutions with different number
$N_s$ of surviving species, also with a few remaining species, or even
just one, as in the black regions.  The differences between the three
graphs are evident from the various sizes and shapes of the coloured
regions. In both ER and scale-free graphs we have cases where only one
species survives (black regions in Fig.~(\ref{fig1b}) and in
Fig.~(\ref{fig1c})) and invades the whole network. However, the black
region is much larger in the phase diagram of the ER random graphs
than in that of the scale-free graphs, where a single species
overcomes all the others only when $p \sim 0.95$ and $\gamma \sim
1.2$. The differences between the phase diagrams of ER and SF graphs
are due to the different degree distribution in the two graphs. In
fact, the diffusion process naturally favours species starting at
poorly connected nodes, since the average number of individuals of
such species that will move to first neighbours is higher than the
average amount of individuals coming from highly connected
nodes. Hence, species starting at poorly--connected nodes have a
higher probability to survive and to invade neighbours. In the ER
graph, these are the few species that survive and invade the graph for
$(p,\gamma)$ in the bottom--right part of the diagram (purple and
black colour), with the competition process involving species starting
at nodes with increasingly large degree, as $p$ decreases.  The same
considerations, based on neighbourhood invasion by species starting at
poorly--connected nodes, hold for scale-free graphs as well, with the
main difference that in a power--law degree distribution the majority
of nodes are poorly--connected, while just a few hubs have a lot of
links. Consequently, species starting at hubs will disappear soon,
while a large number of species tends to survive for a wide range of
$p$ and $\gamma$. This explains why the black region for SF graphs is
much smaller than for ER graphs, and why at any given point
$(p,\gamma)$ of the phase diagram, the number of species $N_s$ on an
ER random graph is always equal or lower than the number of species on
a scale-free graph. The diagram for a regular lattice reported in
Fig.~(\ref{fig1a}) is very similar to that shown in
Fig.~(\ref{fig:rnd_reg_mean_field}).  These results confirm that the
degree distribution of the network plays an important role in the
extinction and survival of species. We have also investigated how the
dynamics of the system depends on other structural properties of the
network, namely the number of nodes, the average node degree, the
clustering coefficient, degree--degree correlations and the presence
of communities. In the following, we use an alternative method to
display the information contained in the phase diagrams. Namely, we
plot the cumulative distribution of the percentage of surviving
species at equilibrium, over all the couples of parameters
$(p,\gamma)$ in the phase diagram. This is useful to compare the phase
diagrams corresponding to different networks.

\begin{figure}[!htbp]
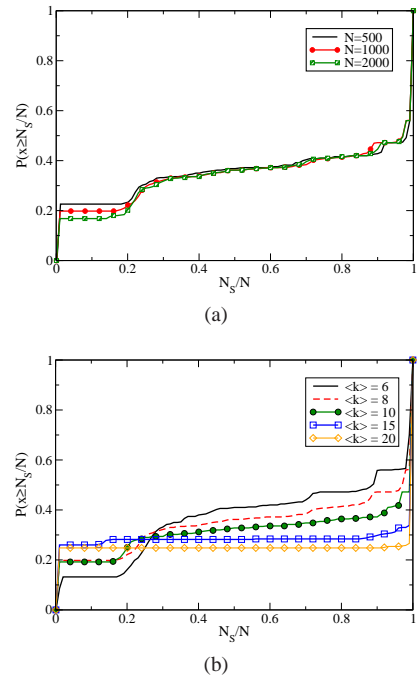

  \centering
  \subfigure[]{
    \includegraphics[scale=0.22]{ER_many_N_8}\label{panel:many_N}
  }\\
  \subfigure[]{
    \includegraphics[scale=0.22]{ER_1000_many_k}\label{panel:many_k}
  }
  \caption{Cumulative distribution of surviving species in
      ER graphs for $0 \le p< 1$ and $1<\gamma\le 6$ (averaged over
      $200$ realizations). Panel (a) $\langle k\rangle = 8$ and
      $N=500$, $N=1000$ and $N=2000$ nodes. Panel (b) $N=1000$ and
      different values of $\langle k \rangle$.}
  \label{fig2}
\end{figure}

In Fig.~(\ref{panel:many_N}) we report the cumulative distribution of
the percentage of surviving species over $(p,\gamma)$ for three ER
random graphs, having average degree $\langle k \rangle=8$ and size
$N=500$, $N=1000$ and $N=2000$, respectively. For the ER random graph
with $N=500$ nodes more than $22\%$ of the $(p,\gamma)$ values cause
the invasion of the network by one species, and $N_s/N=1$ for more
than $45\%$ of the $(p, \gamma)$ values. The percentage of pairs
$(p,\gamma)$ which allow invasion by only one species decreases to
$20\%$ and to $18\%$, respectively for $N=1000$ and $N=2000$, while
there is no appreciable difference among the three curves for
$N_s/N>0.4$. Therefore, for ER random graphs an increase in the
network size producees a slight decrease of the area of the phase
diagram for which we observe invasion by only one species. In
Fig.~(\ref{panel:many_k}) we compare ER random graphs with $N=1000$
nodes and different values of $\langle k\rangle$, namely $6, 8, 10,
15, 20$. As the average degree increases, the shape of the
distribution tends to that of a homogenous network. In particular, for
$\langle k \rangle=20$, the phase diagram is similar to a stepwise
function: only one species survives for $25\%$ of the $(p,\gamma)$
pairs, while all the $N_s=1000$ initial species survive for more than
$60\%$ of the possible $(p,\gamma)$ values.

\begin{figure}[!htb]
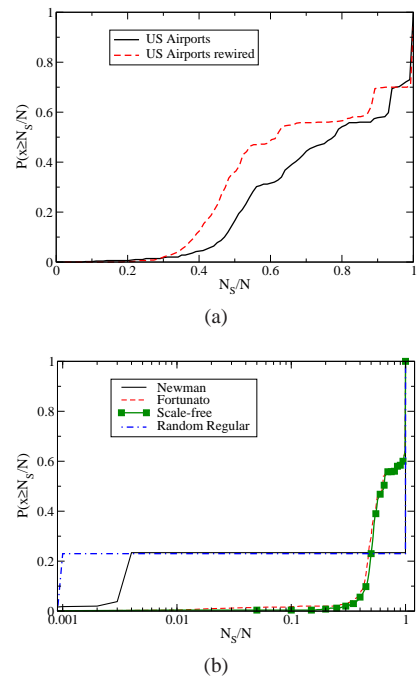

    \centering \subfigure[]{ 
      \includegraphics[scale=0.22]{airport}
      \label{fig3}
    }\\
    \subfigure[]{
      \includegraphics[scale=0.22]{comm_newman_fortunato}
      \label{fig4}
    }
    \caption{Panel~(a): cumulative distribution of surviving
        species for the US Airports network and a random network with
        the same degree distribution.  Panel~(b): cumulative
        distribution of surviving species for the GN benchmark, the
        LFR benchmark, a random regular graph and a scale--free random
        graph. The two benchmark networks have $4$ communities of
        $250$ nodes each.}
  \end{figure}

While ER random graphs are characterised by the size of the network
and by the average degree, networks from the real world usually show
also non-trivial degree--degree correlations and a relatively high
number of triangles. In Fig.~(\ref{fig3}) we report the cumulative
distribution of surviving species at equilibrium for the US Airport
network~\cite{airports}. This network has $N=500$ nodes, representing
airports, $2980$ links, indicating flight connections, a degree
distribution with a power--law tail with an exponent $\sim -1$, a
clustering coefficient $C=0.62$ and disassortative degree--degree
correlations. In the same figure we report the cumulative distribution
of surviving species for a random graph having the same degree
sequence of the US Airports network. The randomisation washes out all
correlations, so that this network has $C = 0.09$ and no
degree--degree correlations. Notice that the US Airports network
allows the survival of a higher percentage of species than its random
counterpart. More than $85\%$ of $(p,\gamma)$ values guarantee the
survival of more than $50\%$ of the species, and for more than $95\%$
of $(p,\gamma)$ values we observe survival of at least $40\%$ of the
species. Conversely, in the random graph only $50\%$ of the species
survive for $60\%$ of $(p,\gamma)$ values. This suggests that even if
two networks have exactly the same degree distribution, the existence
of clustering or degree--degree correlations favours the survival of a
larger number of species.

We have also found that the presence of modules and communities can
affect the dynamics of the system. In Fig.~(\ref{fig4}) we show the
cumulative distribution of the percentage of surviving species for two
standard benchmark networks having a predefined community structure
(the Girvan--Newman's benchmark (GN)~\cite{newman_benchmark} and the
Lancichinetti--Fortunato--Radicchi's benchmark
(LFR)~\cite{fortunato_benchmark}) and for the corresponding random
graphs without any community structure. All the networks have $N=1000$
nodes. The GN benchmark is a regular network with $k = 12$, while the
LFR benchmark is a scale--free network with $P(k) \sim k^{-2}$ and
$\langle k\rangle=6$. The two benchmark networks have $4$ communities
of $250$ nodes each: on average, $90\%$ of the links of each node are
inside its community and the remaining $10\%$ of links point to nodes
outside the community. The distribution of surviving species in the
LFR benchmark is very similar to that of a random scale--free graph:
at least $50\%$ of the species survive for $90\%$ of $(p,\gamma)$
values and more than $70\%$ of the species survive for more than
$45\%$ of $(p, \gamma)$ values. Conversely, the GN benchmark has a
slightly different behaviour compared to the corresponding regular
random graph. In the regular random graph only $1$ species eventually
invades the network for $22\%$ of possible $(p,\gamma)$ values, while
all $N_s$ species survive for $65\%$ of $(p,\gamma)$ pairs. For the GN
benchmark, instead, more than $20\%$ of $(p,\gamma)$ pairs guarantee
the survival of exactly $4$ species. We have checked that in this case
each of the four surviving species is confined into one of the four
communities, and that each node of a given community contains
individuals of only one species. These results indicate that the
existence of communities in the underlying network can affect the
evolution of the system, especially in graphs with homogeneous degree
distributions.

Summing up, we have found that network structure can strongly affect
the dynamics of simple diffusion--competition processes. A central
role in determining the strength of segregation and the number of
surviving species at the equilibrium is played by the degree
distribution. A network with heterogeneous degree distribution
guarantees, for a wider ranges of diffusion and selection parameters,
the survival of a higher number of species compared to the case of a
homogeneous network. In particular, degree heterogeneity helps to
avoid the invasion of the network by only one species.  We have also
investigated the effect of other structural properties, such as the
size of the network, the average degree, the existence of
degree--degree correlations and community structures. In conclusion,
the results confirm that the actual structure of the network has to be
taken seriously into account for the study of competitive processes on
complex topologies, since small differences in the network structure
can produce large differences in the observed dynamics. Our simple
model sheds light on the role of the environment in
diffusion--competition dynamics, and might find useful to explore how
cultures, languages, biotypes and competing populations in general may
survive or get extinct according to the structure of the network they
live in.


\end{document}